\documentclass[twocolumn,twoside]{IEEEtran}
\usepackage[bookmarks=false]{hyperref} 
\usepackage{mathpple}
\usepackage{times}
\usepackage{amsmath}  
\usepackage{amssymb}  
\usepackage{mathrsfs} 
\usepackage{theorem}  
\usepackage{cite}     
\usepackage{comment}  
\usepackage{amsfonts}
\usepackage{verbatim}
\usepackage[dvipsnames,usenames]{color}
\usepackage{enumerate}
\usepackage{graphicx}
\usepackage{latexsym}
\usepackage{ dsfont }
\usepackage{capt-of}
\usepackage{multicol}
\usepackage{hhline}
\usepackage{psfrag}
\usepackage{amsbsy}
\usepackage{tikz}
\usepackage{color}
\usepackage{blkarray} 
\usepackage{graphicx,epsfig}
\usepackage[T1]{fontenc}
\usepackage[sc,osf]{mathpazo}
\usepackage{mathtools}

\newcommand{\be}[1]{\begin{equation}\label{#1}}
\newcommand{\ee}{\end{equation}}

\newcommand{\bc}{\begin{center}}
\newcommand{\ec}{\end{center}}

\newcommand{\cF}{{\cal F}}

\renewcommand{\le}{\leqslant}
\renewcommand{\leq}{\leqslant}
\renewcommand{\ge}{\geqslant}

\newcommand{\Fq}{\smash{\mathbb{F}_{\!q}}}

\newcommand{\C}{\mathbb{C}}

\newcommand{\Cref}[1]{Co\-rol\-la\-ry\,\ref{#1}}

\theoremstyle{plain} \theorembodyfont{\normalfont\slshape}
\newtheorem{thm}{Theorem}
\newenvironment{theorem}{\begin{thm}\hspace*{-1ex}{\bf.}}{\end{thm}}
\newtheorem{prop}[thm]{Proposition$\!$}

\newcounter{cons}
\newenvironment{cons}[1][]{\refstepcounter{cons}\par\medskip\noindent%
   \textbf{Construction~\thecons.#1} \rmfamily}{\medskip}

\newtheorem{fct}[thm]{Fact$\!$}

\newtheorem{lem}[thm]{Lemma$\!$}
\newenvironment{lemma}{\begin{lem}\hspace*{-1ex}{\bf.}}{\end{lem}}
\newtheorem{cor}[thm]{Corollary$\!$}

\newtheorem{defi}{Definition.}

\theorembodyfont{\normalfont}
\newcounter{example}
\newenvironment{example}[1][]{\refstepcounter{example}\par\medskip\noindent%
   \textbf{Example~\theexample.#1} \rmfamily}{\medskip}
\newtheorem{remrk}{Remark$\!$}

\newcounter{algorithm}

\begin{document}

\sloppy

\title{Minimum Storage Regenerating Codes\\ For All Parameters}
 \author{
   \IEEEauthorblockN{
     {\bf Sreechakra Goparaju}\hspace{1.2cm}
     {\bf Arman Fazeli}\hspace{1.2cm}
     {\bf Alexander Vardy}}\\\vspace{0.1cm}
   \IEEEauthorblockA{
     University of California San Diego, La Jolla, CA\,92093, USA\\[0.50ex]
     Email: \{sgoparaju,\,afazelic,\,avardy\}@ucsd.edu}\\
}

\maketitle
\thispagestyle{empty}

\begin{abstract}
Regenerating codes for distributed storage have attracted much 
research interest in the past decade. Such~cod\-es trade the bandwidth 
needed to repair a failed node with the overall amount of data stored 
in the network. Minimum storage regenerating (MSR) codes are an
important class of \emph{optimal} regenerating codes that minimize (first)
the amount of data stored per node and (then) 
the repair bandwidth.
Specifically, an $[n,k,d]$-$(\alpha)$ MSR code $\C$ over $\Fq$ 
is defined as follows. Using such a code $\C$, a file $\cF$ consisting
of $\alpha k$ symbols over~$\Fq$ can be distributed among $n$ nodes,
each storing $\alpha$ symbols, in such a way that: \vspace{-3.00ex}
\begin{list}{}
{
\addtolength{\leftmargin}{-5.50ex}
\addtolength{\rightmargin}{2.50ex}
}
\item\noindent
\begin{itemize}
\item[$\bullet\!\!$]
the file $\cF$ can be recovered by downloading the content of 
any $k$ of the $n$ nodes; and
\item[$\bullet\!\!$]
the content of any failed node can be 
reconstructed~by accessing any $d$ of the remaining $n-1$ nodes
and~down\-loading $\alpha/(d{-}k{+}1)$ symbols from each of these 
nodes.
\end{itemize}
\end{list}
A common practical requirement for regenerating codes is to have
the original file $\cF$ available in {uncoded form}~on~some
$k$ of the $n$ nodes, known as \emph{systematic nodes}. 
In this case, 
several authors~relax the defining 
node-repair condition above, 
requiring the optimal repair bandwidth of $d\alpha/(d{-}k{+}1)$ symbols
for systematic nodes \emph{only}.
We shall call such codes~\emph{sys\-tematic--repair MSR codes}.

Unfortunately, explicit constructions of 
$[n,k,d]$ MSR~codes are known only for
certain special cases: either low rate,~na\-mely $k/n \le 0.5$,
or high repair connectivity, namely $d = n-1$.
Although setting $d = n-1$ minimizes the repair bandwidth,
it may be impractical to connect to \emph{all} the remaining
nodes in order to repair a single failed node. Our main result
in this paper is an explicit construction of systematic-repair
$[n,k,d]$ MSR codes for all possible values of parameters $n,k,d$.
In particular, we construct systematic-repair MSR codes of high
rate $k/n > 0.5$ and low repair connectivity $k \le d \le n-1$.
Such codes were not previously known to exist. 
In order to construct these codes, we solve \emph{simultaneously}
  several repair scenarios, each of which is expressible as an
  interference alignment problem.
Extension of our results beyond systematic repair remains
an open problem. 
\end{abstract}

\section{Introduction}\label{sec:intro}

Distributed storage systems form the backbone for modern cloud computing, large--scale data servers, and peer--to--peer systems. The data in these systems is stored in a redundant fashion --- typically via replication (for instance, Hadoop \cite{Hadoop} and Google file systems \cite{GGL03} adopt a triple replication policy) --- to safeguard data against not--so--infrequently occurring disk failures. An alternative approach to storing data on these systems, which highly reduces the redundancy involved in replication, is to use maximum distance separable (MDS) codes such as Reed--Solomon codes. Though MDS codes are the most space--efficient for a targeted worst--case number of simultaneous node failures, they, unlike repetition codes, incur a high repair bandwidth\footnote{\textcolor{black}{A recent work \cite{GW15} revisits this for the case of Reed--Solomon codes, but we do not go into that here.}} when the system undergoes the repair of a single node failure. A new class of erasure codes, called regenerating codes, was recently defined by Dimakis et al. \cite{DGWWR10} over a set of $n$ nodes, which simultaneously optimizes storage efficiency, worst--case resilience and repair bandwidth for single node failures. These codes follow a trade--off curve which is intuitively evidenced by the contrast between repetition codes and MDS codes: the repair bandwidth decreases as the storage redundancy per node increases.

Formally, a file ${\cal F}$ of size $M$, is said to be stored on a DSS consisting of $n$ nodes, each with a storage capacity of $\alpha$, using an $[n,k,d]$-$(\alpha)$ (or, in short, $[n,k,d]$) regenerating code, if it satisfies two properties: 
\begin{enumerate}
\item[(a)] {\em data recovery}: the file ${\cal F}$ can be recovered using the contents of any $k$ of the $n$ nodes (this property will also be referred to as the {\em MDS property}); and 
\item[(b)] {\em repair property}: the contents of any node can be recovered using the contents of a {\em helper set} of any $d$ other {\em helper} nodes, where each node transmits $\beta$ number of symbols to the replacement node. 
\end{enumerate}
An {\em optimal} $[n,k,d]$ regenerating code achieves the optimal value of total repair bandwidth $\gamma = d\beta$ (minimum repair bandwidth) for a given storage capacity $\alpha$ and $M$. This is given implicitly by the following trade--off:
\begin{eqnarray}
M &=& \sum_{i=0}^{k-1} \min\left\{\alpha, (d-i)\beta\right\}.
\end{eqnarray}

Most of the regenerating codes research (e.g.\cite{SRKR12,RSK11,WD09,CDH09,W09,CJMRS13,CHLM11,PDC11,WTB11,WTB12}) is focussed on the extremal points of this trade--off: MBR and MSR codes. {\em Minimum bandwidth regenerating} (MBR) codes achieve the optimal $\alpha$ when the repair bandwidth equals that of a repetition code. This paper concerns {\em minimum storage regenerating} (MSR) codes, often dubbed as {\em optimal bandwidth MDS codes}, because they are optimal regenerating codes that are {\em also} MDS codes\footnote{To be precise, these are vector MDS codes, i.e., MDS codes over $\mathbb{F}_q^{\alpha}$.}. For these codes, $\alpha = M/k$, and the optimal repair bandwidth is given by:
\begin{eqnarray}
\beta &=& \frac{\alpha}{d-k+1}.
\end{eqnarray}

It is easy to see that the total repair bandwidth $d\beta$ is optimized when the number of helper nodes $d = n-1$. However, it is not always practical to connect to {\em all} the remaining nodes to aid the repair of a failed node. We therefore consider the following question:
{\em Are there constructions of $[n,k,d]$ MSR codes, for $d < n-1$?}

\subsection{Previous Work}
This question has not been wholly unanswered. The first 
MSR code constructions appeared in \cite{SR10,RSK11}, which roughly correspond to the family of parameters $\{n,k,d\}$ with rate $k/n \le 1/2$. The asymptotic existence of MSR codes for all triples $\{n,k,d\}$ was eventually shown in \cite{CJMRS13} using interference alignment techniques developed for a wireless interference channel; these codes achieve optimality as a regenerating code (as well as approach the MSR point) {\em only} when $\alpha \to \infty$, i.e., $\beta/\alpha \to 1/(d-k+1)$, as $M \to \infty$.

MSR codes, being MDS vector codes, can be expressed as a set of $k$ systematic vectors and $n-k$ parity vectors (the corresponding nodes are referred to as systematic and parity nodes, respectively). For the high--rate ($k/n \ge 1/2$) regime, code constructions
were discovered independently
in \cite{CHLM11,PDC11,TWB11,WTB11} for the specific case of $d = n-1$. Of these, the constructions in \cite{CHLM11,PDC11,TWB11} focus on the relaxation of restricting optimal repair to {\em systematic nodes} in the system; we call the corresponding codes {\em systematic--repair MSR codes}. Practical systems usually store information in a systematic format. Parity nodes may fail, but as in the above works, we do {\em not} require optimal bandwidth repair for such nodes (maybe they are less urgent or critical). Clearly, any node can be repaired by reconstructing the whole file, so this covers the node repairability (even if suboptimally). 

\subsection{Contribution \& Outline}
We present the first\footnote{\textcolor{black}{This work was first presented (invited) at the 53rd Annual Allerton 
Conference on Communication, Control, and Computing. A simultaneous result was presented at the same venue by Tamo and En Gad \cite{TE15}. Recently and independently, Rawat et al. \cite{RKV16} have constructed MSR codes which optimally repair all nodes. However, the flavor of their construction, which is not systematic in nature, differs from ours.
}} high--rate finite--$\alpha$ constructions for systematic--repair MSR codes for $d < n-1$. We start by describing in Section \ref{sec:preliminaries} the representative code construction that contains the ideas behind those in \cite{CHLM11,PDC11,TWB11}. Leveraging on this, we present our construction in Section \ref{sec:d<n-1}, but restrict to the case when the helper nodes contain the remaining $k-1$ systematic nodes. This restriction is removed in Section \ref{sec:any_helper_nodes}, thus rounding out the code construction. We conclude with some remarks in Section \ref{sec:conclusion}.

\section{Primer: Code Construction for $d = n-1$}\label{sec:preliminaries}

Let $n=k+r$ denote the number of nodes in the distributed storage system, where each node has the capacity to store a vector of size $\alpha$ over $\mathbb{F}_q$. 
Throughout this paper, we discuss systematic constructions
and assume that the first $k$ nodes are information nodes and store raw information, while the remaining $r$ nodes correspond to the parities. We use the notation ${\bf x}_i,\, i\in [k],$ for the raw information vectors stored in the systematic nodes. The parity nodes are defined by
\begin{align}\label{eq:parity definition}
{\bf x}_{k+i} = \sum_{j=1}^{k} A_{ij} {\bf x}_{j}, \;\;\;\;\;\;\;\;\; i \in [r],
\end{align}
where $A_{ij}$'s are $\alpha\times\alpha$ encoding matrices. The generator matrix of the code is then given by
\begin{align}\label{eq:generator_matrix}
G = \left[ \begin{array}{ccc}
I 		&		&0		\\
		&\ddots 	&		\\
0		&		&I 		\\
A_{1,1}	&\cdots 	&A_{1,k}	\\
\vdots	 	&\ddots	&\vdots	\\
A_{r,1}	&\cdots 	&A_{r,k} 	
\end{array} \right].
\end{align}

In this section, we consider MSR codes where $d=n-1$. In other words, when a single node failure occurs, all the remaining nodes aid in its repair. 
We also restrict our attention to codes that consider failures only of the systematic nodes, and discuss in this section, a construction that underlies the ideas in \cite{TWB11, PDC11} and \cite{CHLM11}. This construction will inform our generalization for the general parameter triple $\{n,k,d\}$ in Section \ref{sec:d<n-1}. 

{\em Remark:} Wang et al. constructed an MSR code for $d = n-1$ in \cite{WTB11} that achieves the optimal repair bandwidth also for parity nodes, albeit at the cost of some other metrics such as the number of symbols read from a node and the complexity of updating parities when systematic data changes. We leave for future the question of whether such a code exists when $d < n-1$.

A commonly adopted strategy in constructing an MSR code is to first guarantee the optimal repair bandwidth property for a single failure (in this case, for a single systematic node failure), and then transform the construction to ensure the MDS property. This is illustrated in Example \ref{ex:423 code} below.

\begin{example}\label{ex:423 code}
Assume $(n,k,d)=(4,2,3)$ and $\alpha=4$. Let the first two nodes ${\bf x}_1$ and ${\bf x}_2$ be the systematic nodes, and let the parity nodes ${\bf x}_3$ and ${\bf x}_4$ be defined as

\[ {\bf x}_3 =\underbrace{ \left( \begin{array}{cccc}
1 	&0	&0	&0	\\
0 	&1	&0	&0	\\
0 	&0	&1	&0	\\
0 	&0	&0	&1	
\end{array} \right)}_{I}{\bf x}_1+
\underbrace{\left( \begin{array}{cccc}
1 	&0	&0	&0	\\
0 	&1	&0	&0	\\
0 	&0	&1	&0	\\
0 	&0	&0	&1	
\end{array} \right)}_{I}{\bf x}_2,\]
\[{\bf x}_4 = \underbrace{\left( \begin{array}{cccc}
0 	&0	&1	&0	\\
0 	&0	&0	&1	\\
1 	&0	&0	&0	\\
0 	&1	&0	&0	
\end{array} \right)}_{P_1}{\bf x}_1+
\underbrace{\left( \begin{array}{cccc}
0 	&1	&0	&0	\\
1 	&0	&0	&0	\\
0 	&0	&0	&1	\\
0 	&0	&1	&0	
\end{array} \right)}_{P_2}{\bf x}_2. 
 \]

Figure~\ref{fig:423 code}.a depicts the component-wise storage in each node. It can be observed that a single failure in either ${\bf x}_1$ or ${\bf x}_2$ can be reconstructed by downloading $\alpha/2=2$ elements from each of the remaining $d=3$ nodes. However, the data is not recoverable if both ${\bf x}_1$ and ${\bf x}_2$ fail and hence, the code is not MDS. To overcome this problem, we associate a coefficient $\lambda$ with $P_2$ such that $\small{\begin{pmatrix} I & I\\  P_1 & \lambda P_2  \end{pmatrix}}$ is non-singular. Note that,
\begin{align*}
\begin{vmatrix} I &\hspace{-1.5ex} I\\
  P_1 & \hspace{-1.5ex}\lambda P_2  \end{vmatrix} \hspace{-.5ex}=\hspace{-.3ex} \det(\lambda P_2 - P_1)\hspace{-.5ex}=\hspace{-.5ex} 
\begin{vmatrix} 
0 		&\hspace{-.7ex}\lambda	&\hspace{-.7ex}-1		&\hspace{-.7ex}0	\\
\lambda 	&\hspace{-.7ex}0		&\hspace{-.7ex}0		&\hspace{-.7ex}-1	\\
-1 		&\hspace{-.7ex}0		&\hspace{-.7ex}0		&\hspace{-.7ex}\lambda	\\
0 		&\hspace{-.7ex}-1		&\hspace{-.7ex}\lambda	&\hspace{-.7ex}0	
\end{vmatrix}\hspace{-.5ex}=\hspace{-.5ex} (\lambda^2 -1)^2,
\end{align*}
which is non-zero\footnote{In general, if $A,B,C,$ and $D$ are nonsingular $\alpha\times\alpha$ matrices, then $\det\big({\begin{bmatrix}A &B\\ C &\lambda D\end{bmatrix}}\big)$ is given by $\det(A)\det(\lambda D - CA^{-1}B)$, which is a polynomial of degree at most $\alpha$ in $\lambda$. If the field size is large enough, \emph{i.e.} $q>\alpha$, one can always find a value for $\lambda$ so that the $2\times 2$ block matrix becomes non-singular as well. The same approach can be used to prove Lemma~\ref{lemma:MDS property2}. } if $q=5$ and $\lambda = 2$. Figure~\ref{fig:423 code}.b shows the component-wise storage for the resulting MSR code. 

\begin{figure}[t]
	\centering
	\includegraphics[scale=.5]{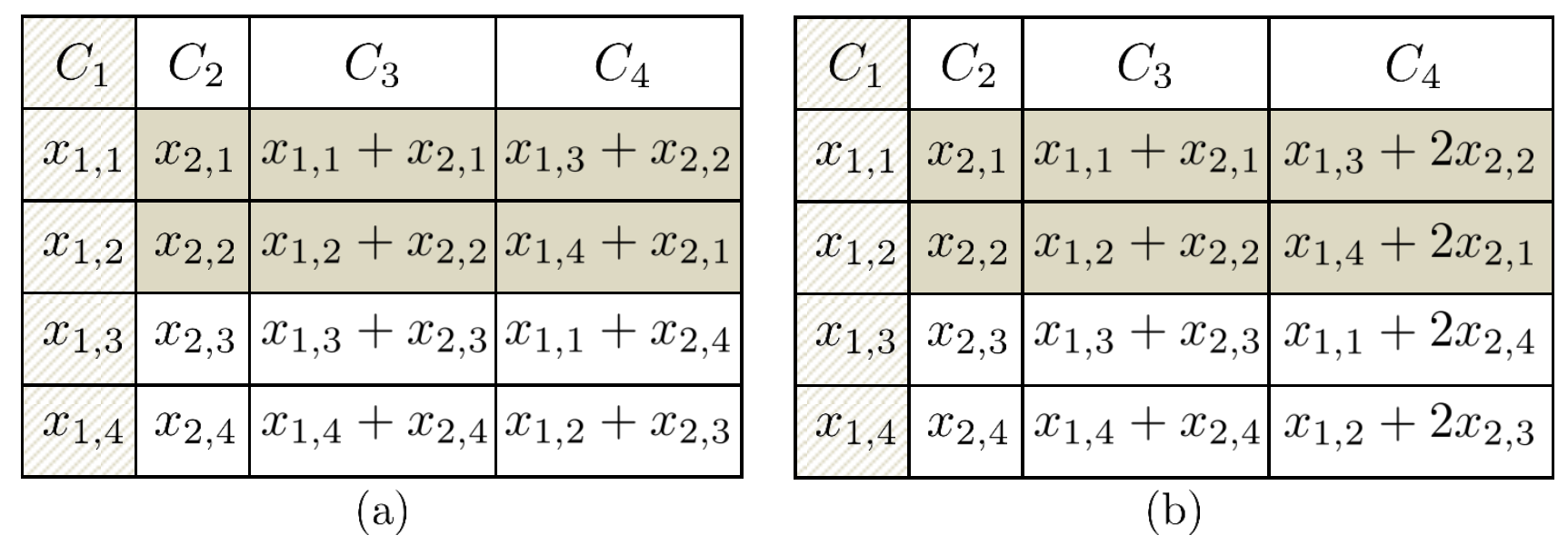}
	\caption{{(a) Component wise storage in a $(4,2,3)$ binary array code with optimal repair bandwidth for a single systematic node failure, described by $({\bf x}_1, {\bf x}_2, {\bf x}_1 + {\bf x}_2, P_1{\bf x}_1 + P_2{\bf x}_2)$; (b) A $(4,2,3)$ MSR code in $\mathbb{F}_5$ described by $({\bf x}_1, {\bf x}_2, {\bf x}_1 + {\bf x}_2, P_1{\bf x}_1 +2 P_2{\bf x}_2)$. In both cases, gray cells are accessed to rebuild $C_1$. 
	\label{fig:423 code}}}
\end{figure}
\end{example}

Construction \ref{cons:d_n-1_systematic} generalizes the construction given in Example \ref{ex:423 code} for an $(n,k,n-1)$ MSR code. Note that any MSR code construction must specify both the generator matrix of the code as well as the optimal bandwidth repair strategy that is implemented on the code.

\begin{cons}\label{cons:d_n-1_systematic} Let $\alpha = r^k$ and label the $\alpha$ elements $[0:r^k-1]$ by $r$-ary vectors in $\mathbb{Z}_r^k$.  Define permutation $f_j^{\ell}$ on $[0:r^k-1]$ as follows:
\begin{eqnarray*}
\begin{array}{cccl}
f_j^{\ell}:&\mathbb{Z}_r^k &\to& \mathbb{Z}_r^k\\
&v &\mapsto& v + \ell e_j,
\end{array}
\end{eqnarray*} 
for $j\in [k]$ and $\ell \in [0:r-1] := \{0,1,\ldots,r-1\}$, where $\{e_1,e_2,\ldots,e_k\}$ is the standard vector basis for $\mathbb{Z}_r^k$. 
The mapping $f_j^{\ell}$ is bijective, and therefore, corresponds to a permutation on $[0:r^k-1]$.
Let $P_{\ell,j}$ be the $\alpha\times\alpha$ matrix corresponding to the permutation $f_j^{\ell}$, that is, $P_{\ell,j}\,{\bf x} = {\bf y}$, where ${\bf x}$, ${\bf y} \in \mathbb{F}_q^{\alpha}$, and ${\bf x}(v) = {\bf y}(f_j^{\ell}(v))$. In other words, $P_{\ell,j}$ scrambles the elements of a vector according to the permutation $f_j^{\ell}$. (Notice that $P_{0,j} = I_{\alpha}$.)
\begin{enumerate}
\item {\em MSR Code}: The generator matrix of the code is given by (\ref{eq:generator_matrix}), where $A_{i,j} = \lambda_{i,j}P_{i-1,j}$, $i \in [r]$ and $j \in [k]$. 
The non-zero coefficients $\lambda_{i,j}\in \mathbb{F}_q$ will be defined in Section \ref{sec:MDS} to ensure the MDS property.

\item {\em Repair Strategy}: Let $Y_j=\{v\in[0,r^k-1]:v \cdot e_j=0 \}$ denote a subset of $[0:r^k-1]$. $Y_j$ can be interpreted as those elements in $[0:\alpha-1]$ whose label representation in $\mathbb{Z}_r^k$ have a $0$ in their $j^{\text{th}}$ coordinate. If systematic node $j$ fails, it is repaired by accessing the elements corresponding to $Y_j$ from each of the remaining nodes, i.e., by accessing ${\bf x}_i(v)$, where $v \in Y_j$ and $j \neq i \in [n]$.
\end{enumerate}
\end{cons}

Construction \ref{cons:d_n-1_systematic} is obtained by first constructing an $[n,k]$ array code\footnote{By an $[n,k]$ array code, we mean a set of $k$ systematic vectors, and $n-k$ parity vectors defined according to (\ref{eq:parity definition}), which may or may not satisfy any properties.} (Section \ref{sec:IA}) which guarantees the optimal bandwidth repair for a single systematic node failure. The array code is then transformed (Section \ref{sec:MDS}) to an MDS array code (and thereby, a systematic--repair MSR code) by transforming the encoding matrices of the parity nodes, while retaining the repair property.

\subsection{Repair Property: Interference Alignment}\label{sec:IA}
The optimal repair bandwidth property of an $[n,k,n-1]$ MSR code can be viewed as a {\em signal interference} problem: the objective is to retrieve the desired signal --- the contents of the failed systematic node, say, ${\bf x}_i$ --- which, in the repair data downloaded from the remaining nodes, is interfered by partial contents of the remaining systematic nodes, ${\bf x}_j$, where $i \neq j \in [n]$. The solution, turns out to be an {\em interference alignment} strategy, where the repair data associated with the interfering systematic data is aligned, so as to minimize the interference. This is crystallized in the following lemma\footnote{This result is known and has been used in several papers on MSR codes, but we state and prove it for completeness.}.

\begin{lemma}\label{lemma:interference alignment_n-1}
Let ${\bf x}_i$, $i \in [k]$, be the failed systematic node. For an $[n,k,n-1]$ MSR code, the set of $d = n-1$ helper nodes is given by ${\cal D} = \{{\bf x}_j \,|\, j \in [n]\backslash \{i\}\}$. To recover the contents of the failed systematic node with the optimal repair bandwidth, it is necessary and sufficient to find $n-1$ (repair) matrices denoted by 
 $\{S_j^i\in \mathbb{F}_q^{\alpha/r \times \alpha}\,|\,j \in [n]\backslash \{i\}\}$, where $r = n-k$, such that, for $j \in [k], j \neq i$, the following two conditions are satisfied:
 
\noindent (a) signal recovery:
\begin{eqnarray}\label{eq:signal-recovery}
\mathsf{rank}\left(\left(\begin{array}{c}
S_{k+1}^iA_{1,i}	\\
S_{k+2}^iA_{2,i}	\\
\vdots 	\\
S_{k+r}^iA_{r,i}	
\end{array}\right)\right) &=& \alpha,
\end{eqnarray}
(b) interference alignment:
\begin{eqnarray}\label{eq:IAcondition}
\mathsf{rank}\left(\left(\begin{array}{c}
S_j^i	\\
S_{k+1}^iA_{1,j}	\\
\vdots 	\\
S_{k+r}^iA_{r,j}	
\end{array}\right)\right) &=& \frac{\alpha}{r}.
\end{eqnarray}

Stated otherwise, to optimally repair ${\bf x}_i$, it is necessary and sufficient to find $n-1$ (repair) subspaces of dimension $\alpha/r$, denoted\footnote{Whenever this lemma is referenced, we use the subspace and matrix notation interchangeably as some proofs or expressions are clearer in one of the formats. We accordingly overload the notation $S_j^i$ to refer to both the matrix and the subspace spanned by the row vectors of the matrix.} by $\{S_j^i\,|\,j \in [n]\backslash \{i\}\}$, where $r = n-k$, such that, for $j \in [k], j \neq i$, the following two conditions are satisfied:

\noindent (a) signal recovery:
\begin{eqnarray}\label{eq:signal-recovery2}
S_{k+1}^iA_{1,i} \oplus \cdots \oplus S_{k+r}^iA_{r,i} &\backsimeq& \mathbb{F}_q^{\alpha},
\end{eqnarray}
(b) interference alignment:
\begin{eqnarray}\label{eq:IAcondition2}
S_j^i &\backsimeq& S_{k+s}^iA_{s,j},\,\,\forall \,s \in [r],
\end{eqnarray}
where $\backsimeq$ denotes equality of subspaces, $SA$ is the subspace obtained by operating the subspace $S$ by the matrix $A$, and $\oplus$ denotes the subspace sum.
\end{lemma}

For completeness, we provide a proof for Lemma \ref{lemma:interference alignment_n-1} in Appendix \ref{app:lemmaIA}. Lemma \ref{lemma:interference alignment_general} generalizes Lemma \ref{lemma:interference alignment_n-1} when the number of helper nodes $d < n-1$. This will be used later in Section \ref{sec:d<n-1}.

\begin{lemma}\label{lemma:interference alignment_general}
{\em (Corollary of Lemma \ref{lemma:interference alignment_n-1}.)}
In general, for an $[n,k,d]$ MSR code, if the set of $d=(k-1)+t<n-1$ helper nodes is given by ${\cal D} = \{{\bf x}_j \,|\, j \in {\cal J} = [k]\backslash \{i\}\cup\{b_1,\cdots,b_t\}\}$ (where $b_i \in \{k+1,\ldots,n\}$ denote the $t$ parity nodes in the helper set), it is necessary and sufficient to find $d$ (repair) subspaces of dimension $\alpha/t$ denoted by $\{S_j^i\,|\,j \in {\cal J}\}$,
such that, for $j \in [k], j \neq i$, the following two conditions are satisfied: 
\begin{eqnarray}
S_{b_1}^iA_{b_1-k,i} \oplus \cdots \oplus S_{b_t}^iA_{b_t-k,i} &\backsimeq& \mathbb{F}_q^{\alpha},
\end{eqnarray}
\begin{eqnarray}
S_j^i &\backsimeq& S_{b_s}^iA_{b_s-k,j},\,\,\forall \,s \in [t].
\end{eqnarray}
\end{lemma}

The optimal repair property of Construction \ref{cons:d_n-1_systematic} can now be justified. 

\begin{figure}[t]
	\centering
	\includegraphics[scale=.33]{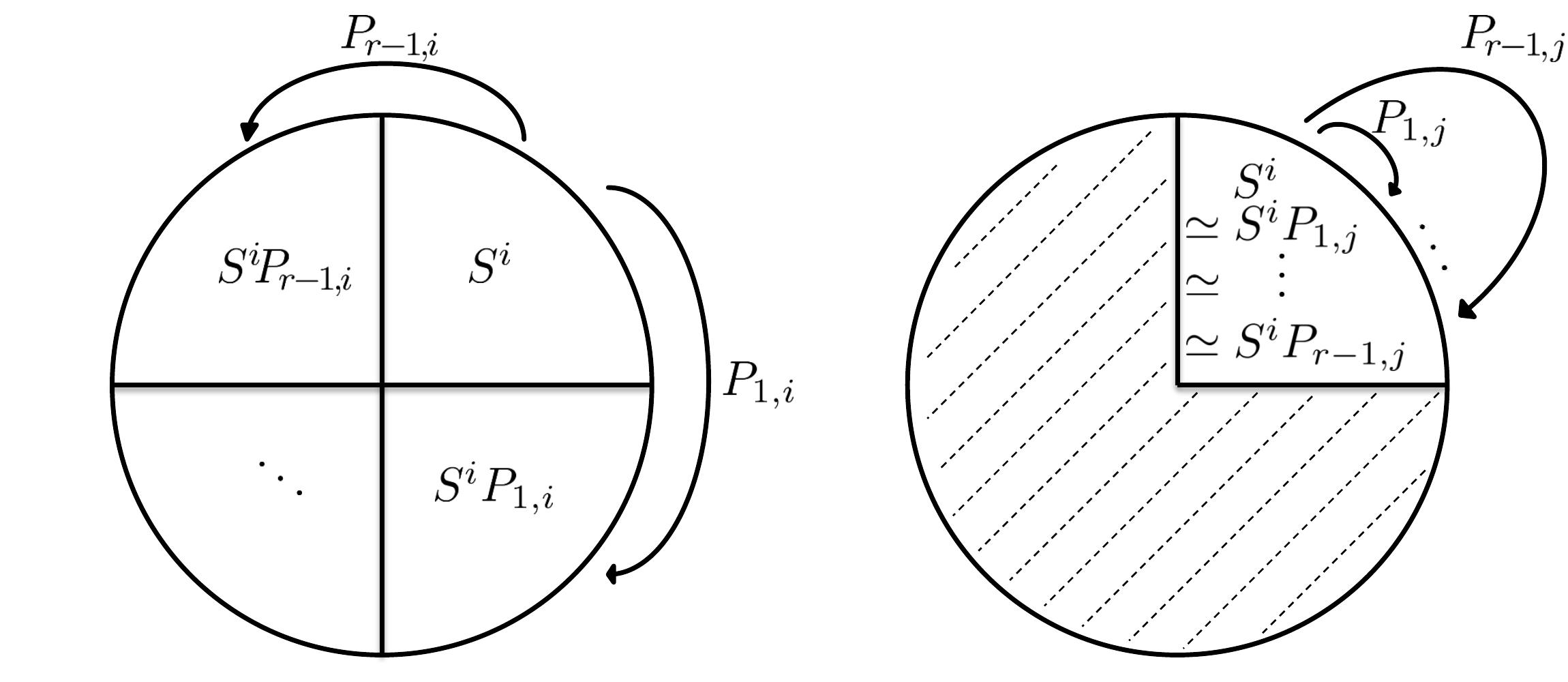}
	\caption{Visualization of Lemma \ref{lemma:interference alignment_n-1}(a)[left], and Lemma \ref{lemma:interference alignment_n-1}(b)[right] to justify repair optimality in Construction~\ref{cons:d_n-1_systematic}.}
\end{figure}

\begin{lemma}\label{lemma:optimal-repair-d_n-1}
The repair strategy in Construction \ref{cons:d_n-1_systematic} is optimal with respect to repair bandwidth.
\end{lemma}
\begin{IEEEproof}
Define $S_j^i\triangleq S^i\triangleq Y_i, j \neq i$. Notice that the rank of subspace $S^i$ is $r^{k-1}=\alpha/r$. Per definition, the permutation $P_{\ell,i}$ maps $Y_i$ to $Y_i + \ell e_i = \{v\in[0,r^k-1]:v \cdot e_i=\ell \}$. This implies that for any distinct $\ell, \ell' \in [0:r-1]$, the intersection $S^iP_{\ell,i} \cap S^iP_{\ell',i}$ contains only the all-zero vector. Thus the subspaces: $S^i, S^iP_{1,i}, \ldots, S^iP_{r-1,i}$, span the space $\mathbb{F}_q^{\alpha}$ ($\alpha = r^k$) and the signal recovery condition(s) in Lemma \ref{lemma:interference alignment_n-1} are satisfied. Furthermore, applying a permutation $P_{\ell,j}$ corresponding to a different coordinate $j \neq i$ maps $Y_i$ to itself.
This validates the interference alignment condition(s) in Lemma \ref{lemma:interference alignment_n-1}. Finally, note that the two conditions continue to be satisfied when replacing the permutations $P_{i-1,j}$ with any scaled versions $A_{i,j} = \lambda_{i,j}P_{i-1,j}$, because the scaling of the basis vectors does not change the relevant subspaces and thereby does not affect the conditions in Lemma \ref{lemma:interference alignment_n-1}.
\end{IEEEproof}

\subsection{MDS Property}\label{sec:MDS}
This second step relies on the following two lemmas, the proofs of which are left to the reader.
\begin{lemma}\label{lemma:MDS property1}
Let $B$ denote the parity part of the generator matrix for an $[n,k]$ array code denoted by $\mathbb{C}$, where
\[ B = \left[ \begin{array}{ccc}
B_{1,1}	&\cdots	&B_{1,k}	\\
\vdots		&\ddots 	&\vdots	\\
B_{r,1}	&\cdots	&B_{r,k}	
\end{array} \right].\]
Given that $B_{i,j}$ is non-singular for all $i,j$, then $\mathbb{C}$ is an MDS array code if and only if any square sub-block-matrix $B'$ of $B$ is also non-singular, where 
\[ B' = \left[ \begin{array}{ccc}
B_{i_1,j_1}	&\cdots	&B_{i_1,j_t}	\\
\vdots		&\ddots 	&\vdots	\\
B_{i_t,j_1}	&\cdots	&B_{i_t,j_t}	
\end{array} \right],\;\;\]
for some $\{i_1,\cdots,i_t\}\subset [r], \{j_1,\cdots,j_t\}\subset [k]$.
\end{lemma}

\begin{lemma}\label{lemma:MDS property2}
Let $B$ denote the $r\alpha\times k\alpha$ matrix associated with the parity part of the generator matrix for an $[n,k]$ array code, as defined in Lemma~\ref{lemma:MDS property1}. 
Given that $B_{i,j}$ is non-singular for all $i \in [r]$, $j \in [k]$, and the field size $q$ is large enough, there exist coefficients $\lambda_{i,j}\in \mathbb{F}_q$, 
such that all square sub-block-matrices of $A$  are non-singular, where
\[ A = \left[ \begin{array}{ccc}
\lambda_{1,1}B_{1,1}	&\cdots	&\lambda_{1,k}B_{1,k}	\\
\vdots		&\ddots 	&\vdots	\\
\lambda_{r,1}B_{r,1}	&\cdots	&\lambda_{r,k}B_{r,k}	
\end{array} \right].\]
In other words, any parity generator matrix $B$ for an $[n,k]$ array code with non-singular encoding matrices can be transformed into a parity generator matrix $A$ for an $[n,k]$ MDS array code by multiplying the encoding matrices with appropriate scalar coefficients.
\end{lemma}

\begin{IEEEproof}[\textcolor{black}{Proof Sketch}] To obtain a valid set of $\lambda_{i,j}$'s, one may first sort the pairs $(i,j)$ with respect to $i+j$ increasingly, and then recursively choose a value for each $\lambda_{i,j}$ such that all sub-block-matrices with $\lambda_{i,j}A_{i,j}$ on their bottom right corner become non-singular. It suffices to have the field size $q$ greater than the number of such sub-block-matrices at any step multiplied by $\alpha$;
\begin{eqnarray*}
|\mathbb{F}|>q_{\text{\tiny{MDS}}}=\alpha\max_{t}\left\{\displaystyle\hspace{-.5ex}\binom{\hspace{-.2ex}n-k-1\hspace{-.2ex}}{t}\hspace{-.5ex}\times\hspace{-.5ex}\binom{k-1}{t}\bigg|t\in[k]\right\}\hspace{-.6ex}. \hspace{0.55cm}
\end{eqnarray*}
\end{IEEEproof}

\section{Code Construction for Restricted Helper Set}\label{sec:d<n-1}
We now move to the construction of $[n,k,d]$ systematic--repair MSR codes for any $n$, $k$, and $d$, where $k+1\leq d \leq n-1$. 
In this section, we start with the restricted case when the helper set $\mathcal{D}$ includes all remaining $k-1$ systematic nodes. Let us begin with an example.
\begin{example}\label{example:k+3_k_k+1}
Let us look at the case when $[n,k,d]= [k+3,k,k+1]$ for $k\in\mathbb{N}$. Given a failure at the systematic node $i$, we are interested in repairing it by downloading $\frac{\alpha}{d-k+1} = \frac{\alpha}{2}$ symbols from each node in the helper set $\mathcal{D}_i$. Let us assume that $\mathcal{D}_i$ includes all of the remaining $k-1$ systematic nodes. Hence, there are $\binom{3}{2}=3$ different ways to choose ${\cal D}_i$ depending on which two parity nodes are included in it. Let us use an indicator $a\in[3]$ to differentiate between these scenarios, and denote the helper set for each scenario by $\mathcal{D}_{i,a}$.

\emph{Construction.} Let $\alpha=2^{3k}$ and label the $\alpha$ elements $[0:2^{3k}-1]$ by binary vectors in $\mathbb{Z}_2^{3k}$. Define permutation $f_{j}^{\ell}$ on $[0:2^{3k}-1]$ as follows:
\begin{align*}
&f_{j}^\ell : 	&\hspace{-7ex}\mathbb{Z}_2^{3k} 	&\hspace{2ex}\to 		&\hspace{-9ex}\mathbb{Z}_2^{3k}\hspace{3.3ex}\\
&		&\hspace{-7ex}v 				&\hspace{2ex}\mapsto	&\hspace{-9ex}v+\ell e_i,
\end{align*}
for $j\in[3k]$ and $\ell\in\{0,1\}$, where $\{e_1,e_2,\cdots,e_{3k}\}$ is the standard vector basis for $\mathbb{Z}_2^{3k}$. The mapping $f_j^{\ell}$ is again bijective and therefore corresponds to a permutation on $[0:2^{3k}-1]$. As before, let $P_{\ell,j}$ be the $\alpha\times\alpha$ matrix corresponding to the permutation $f_j^{\ell}$, that is, $P_{\ell,j}{\bf x} = {\bf y}$, where ${\bf x}, {\bf y} \in \Fq^{\alpha}$, and ${\bf x}(v) = {\bf y}(f_j^{\ell}(v))$. (Notice again that $P_{0,j} = I_{\alpha}$.)

\begin{enumerate}
\item {\em MSR Code}: The generator matrix of the code is given by
\[ G = \left[ \begin{array}{ccc}
I 		&		&0		\\
		&\ddots 	&		\\
0		&		&I 		\\
A_{1,1}	&\cdots 	&A_{1,k}	\\
A_{2,1}	&\cdots 	&A_{2,k}	\\
A_{3,1}	&\cdots 	&A_{3,k} 	
\end{array} \right],\]
where
\begin{align}\label{eq:example_2_construction}
&A_{1,j}=\lambda_{1,j}\hspace{1ex}\times\hspace{1ex} P_{0,3j-2} \hspace{1ex}\times \hspace{1ex} P_{0,3j-1} \hspace{1ex}\times I_{\alpha},\nonumber\\\nonumber
&A_{2,j}=\lambda_{2,j}\hspace{1ex}\times \hspace{1ex}P_{1,3j-2} \hspace{1ex}\times  \hspace{1ex}I_{\alpha}   \hspace{5.15ex}\times P_{0,3j},\\
&A_{3,j}=\lambda_{3,j}\hspace{1ex}\times \hspace{1ex} I_{\alpha} \hspace{5.15ex}\times \hspace{1ex}P_{1,3j-1} \hspace{1ex}\times P_{1,3j},
\end{align}
for $j\in [k]$. The non-zero coefficients $\lambda_{i,j}\in \Fq$ are again selected according to the discussion in Section \ref{sec:MDS} to establish the MDS property. 
\item {\em Repair Strategy via  $\mathcal{D}_{i,1} = \{{\bf x}_j | j\in [k+3], j\neq i, k+3 \}$}: 

Let $Y_{i,1}=\{v\in[0,2^{3k}-1]:v \cdot e_{3i-2}=0 \}$ denote a subset of $[0:2^{3k}-1]$. $Y_{i,1}$ can be interpreted as those elements in $[0:2^{3k}-1]$ whose label representation in $\mathbb{Z}_2^{3k}$ have a $0$ in their $(3i-2)^{\text{th}}$ coordinate. If systematic node $i$ fails, it can be repaired by accessing the elements corresponding to $Y_{i,1}$ from each of the helper nodes, i.e., by accessing ${\bf x}_j(v)$, where $v \in Y_{i,1}$ and $j \in \{1,2,\cdots,i-1,i+1,\cdots,k,k+1,k+2\}$.
\item {\em Repair Strategy via $\mathcal{D}_{i,2} = \{{\bf x}_j | j\in [k+3], j\neq i, k+2 \}$}: 

Similarly, let $Y_{i,2}=\{v\in[0,2^{3k}-1]:v \cdot e_{3i-1}=0 \}$. If systematic node $i$ fails, it can be repaired by accessing ${\bf x}_j(v)$, where $v \in Y_{i,2}$ and $j \in \{1,2,\cdots,i-1,i+1,\cdots,k,k+1,k+3\}$. 
\item {\em Repair Strategy via $\mathcal{D}_{i,3} = \{{\bf x}_j | j\in [k+3], j\neq i, k+1 \}$}: 

Finally, let $Y_{i,3}=\{v\in[0,2^{3k}-1]:v \cdot e_{3i}=0 \}$ denote the location of the elements that have to get accessed if the systematic node $i$ fails, i.e., node $i$ can be repaired by accessing ${\bf x}_j(v)$, where $v \in Y_{i,3}$ and $j \in \{1,2,\cdots,i-1,i+1,\cdots,k,k+2,k+3\}$. \\
\end{enumerate}
\begin{IEEEproof}[Justification of the repair strategy]
Let ${\bf x}_i$, $i\in[k]$, be the failed systematic node. Define $Q_{u,v}=A_{u,v}\lambda_{u,v}^{-1}$, $u\in[3], v\in[k]$, which is a product of multiple permutation matrices, and hence can be viewed as a permutation matrix itself. In order to justify the repair strategy, it suffices to define the proper subspaces $S_j^i$ that fulfill the two interference alignment conditions in Lemma \ref{lemma:interference alignment_general}. Let $U_{i,a}$ be the complimentary subset of $Y_{i,a}$ in $\mathbb{Z}_2^{3k}$, i.e., 
\begin{align*}
&U_{i,1}=\{v\in[0,2^{3k}-1]:v \cdot e_{3i-2}=1 \},\\
&U_{i,2}=\{v\in[0,2^{3k}-1]:v \cdot e_{3i-1}=1 \},\\
&U_{i,3}=\{v\in[0,2^{3k}-1]:v \cdot e_{3i}=1 \}.
\end{align*}
Given the code construction in (\ref{eq:example_2_construction}), we can verify that
\begin{align}\label{eq:example2_justification}
Y_{i,1}Q_{1,i} &= Y_{i,1},&	Y_{i,2}Q_{1,i} &= Y_{i,2},	&Y_{i,3}Q_{1,i} &= Y_{i,3},\nonumber\\
U_{i,1}Q_{1,i} &= U_{i,1},&	U_{i,2}Q_{1,i} &= U_{i,2},	&U_{i,3}Q_{1,i} &= U_{i,3},\nonumber\\
Y_{i,1}Q_{2,i} &= U_{i,1},&	Y_{i,2}Q_{2,i} &= Y_{i,2},	&Y_{i,3}Q_{2,i} &= Y_{i,3},\nonumber\\
U_{i,1}Q_{2,i} &= Y_{i,1},&	U_{i,2}Q_{2,i} &= U_{i,2},	&U_{i,3}Q_{2,i} &= U_{i,3},\nonumber\\
Y_{i,1}Q_{3,i} &= Y_{i,1},&	Y_{i,2}Q_{3,i} &= U_{i,2},	&Y_{i,3}Q_{3,i} &= U_{i,3},\nonumber\\
U_{i,1}Q_{3,i} &= U_{i,1},&	U_{i,2}Q_{3,i} &= Y_{i,2},	&U_{i,3}Q_{3,i} &= Y_{i,3},\nonumber\\
 \hspace{-2ex}\text{and,}  & & & & \nonumber\\
Y_{i',u}Q_{i,j} &= Y_{i',u}\hspace{.5ex},\hspace{-.5ex}&		U_{i',u}Q_{i,j} &= U_{i',u}\hspace{.5ex},\hspace{-.5ex}		&\text{for } &i'\neq i.
\end{align}
Now we define subspaces $S_{j,a}^i \triangleq S_{a}^i \triangleq Y_{i,a}, j\neq i, a\in [3]$. Let us for simplicity assume $a=1$. The other scenarios follow the proof similarly. Based on (\ref{eq:example2_justification}), we observe that the permutation $Q_{2,i}$ maps the basis $Y_{i,1}$ to its complementary subset $U_{i,1}$ and vice versa, while $Q_{1,i}$ preserves both of them. Hence,
\begin{align*}
\mathsf{rank}\left( \left( \begin{array}{c}
S_1^iQ_{1,i}	\\
S_1^iQ_{2,i}	
\end{array} \right) \right) = 
\mathsf{rank}\left( \left( \begin{array}{c}
Y_{i,1}	\\
U_{i,1}
\end{array} \right) \right) =\alpha.
\end{align*}
Furthermore, $Y_{i,1}$ remains unchanged under any other permutation $Q_{t,j}, j\neq i$, and hence
\begin{align*}
\mathsf{rank}\left( \left( \begin{array}{c}
S_1^i	\\
S_1^iQ_{1,i'}	\\
S_1^iQ_{2,i'}	
\end{array} \right) \right) = 
\mathsf{rank}\left( \left( \begin{array}{c}
Y_{i,1}	\\
Y_{i,1}	\\
Y_{i,1}	
\end{array} \right) \right) =\frac{\alpha}{2}.
\end{align*}
\end{IEEEproof}
\end{example}
The key element in the construction is to satisfy the two requirements in Lemma \ref{lemma:interference alignment_general} for any systematic failure and any such helper set $\mathcal{D}$. Let $\rho=d-k+1$ denote the number of parity nodes in the helper set of size $d$. There are $\binom{r}{\rho}$ different ways to choose $\rho$ parity nodes during the repair. Let us label these cases with numbers $a\in [\binom{r}{\rho}]$, and set $\mathcal{R}_a$ to be the subset of parity nodes corresponding to case $a$. 

Assume that $\mathcal{R}_a=\{{\bf x}_{k+d_1^{(a)}},{\bf x}_{k+d_2^{(a)}},\cdots,{\bf x}_{k+d_{\rho}^{(a)}}\}$ is the ordered representations, where $\{d_1^{(a)},\cdots,d_{\rho}^{(a)}\} \subset [r]$. Finally, define $r$-ary vectors ${\omega}_{a}$ for $a\in[\binom{r}{\rho}]$ as
$$\omega_a(i) = \left\{
	\begin{array}{lll}
		t-1  && \mbox{if } \exists t:  i=d_t^{(a)},\\
		0 && \mbox{otherwise.}
	\end{array}
\right. $$

\begin{cons}\label{cons:fixed_d}
Let $\alpha =\rho^{k\binom{r}{\rho}}$ and label the $\alpha$ elements $[0:\alpha-1]$ by $\rho$-ary vectors in $\mathbb{Z}_{\rho}^{k\binom{r}{\rho}}$.  Define permutation $f_j^{\ell}$ on $[0:\alpha-1]$ as follows:
\begin{eqnarray*}
\begin{array}{cccl}
f_j^{\ell}:&\mathbb{Z}_{\rho}^{k\binom{r}{\rho}} &\to& \mathbb{Z}_{\rho}^{k\binom{r}{\rho}}\\
&v &\mapsto& v + \ell e_j,
\end{array}
\end{eqnarray*} 
for $j \in [k\binom{r}{\rho}]$ and $\ell \in [0:\rho-1]$, where $\{e_1,\cdots,e_{k\binom{r}{\rho}}\}$ is the standard vector basis of $\mathbb{Z}_{\rho}^{k\binom{r}{\rho}}$. Let $P_{\ell,j}$ be the $\alpha\times\alpha$ matrix corresponding to the permutation $f_j^{\ell}$.
\begin{enumerate}
\item {\em MSR Code}: The generator matrix of the $[n,k,d]$ code is given by (\ref{eq:generator_matrix}), where 
\begin{align*}
A_{i,j}=\lambda_{i,j}\prod_{a\in [\binom{r}{\rho}]}
P_{w_{a}(i),\hspace{.2ex}a+(j-1)\binom{r}{\rho}}, &\hspace{2ex}\text{for } j\in[k],i\in[r].
\end{align*}
The non-zero coefficients $\lambda_{i,j}\in \mathbb{F}_q$ are defined according to Section \ref{sec:MDS} to ensure the MDS property; and later will be modified again in Section \ref{sec:any_helper_nodes}.
\item {\em Repair Strategy}: Let ${\cal R}_a$ correspond to the parity subset of the helper set $\mathcal{D}$. Define $Y_{j,a}\subset[0:\alpha-1]$  as $\{x\in[0,\alpha-1]:x \cdot e_{a+(j-1)\binom{r}{\rho}}=0 \}$. If systematic node $j$ fails, it is repaired by accessing the elements corresponding to $Y_{j,a}$ from helper nodes, i.e., by accessing ${\bf x}_i(v)$, where $i\in\mathcal{D}$, and $v \in Y_{j,a}$.
\end{enumerate}
\end{cons}

\begin{lemma}\label{thm:fixed_d} 
The repair strategy in {\em Construction \ref{cons:fixed_d}} is optimal with respect to repair bandwidth. 
\end{lemma}

\begin{figure}[t]
\centering
\includegraphics[scale=0.5]{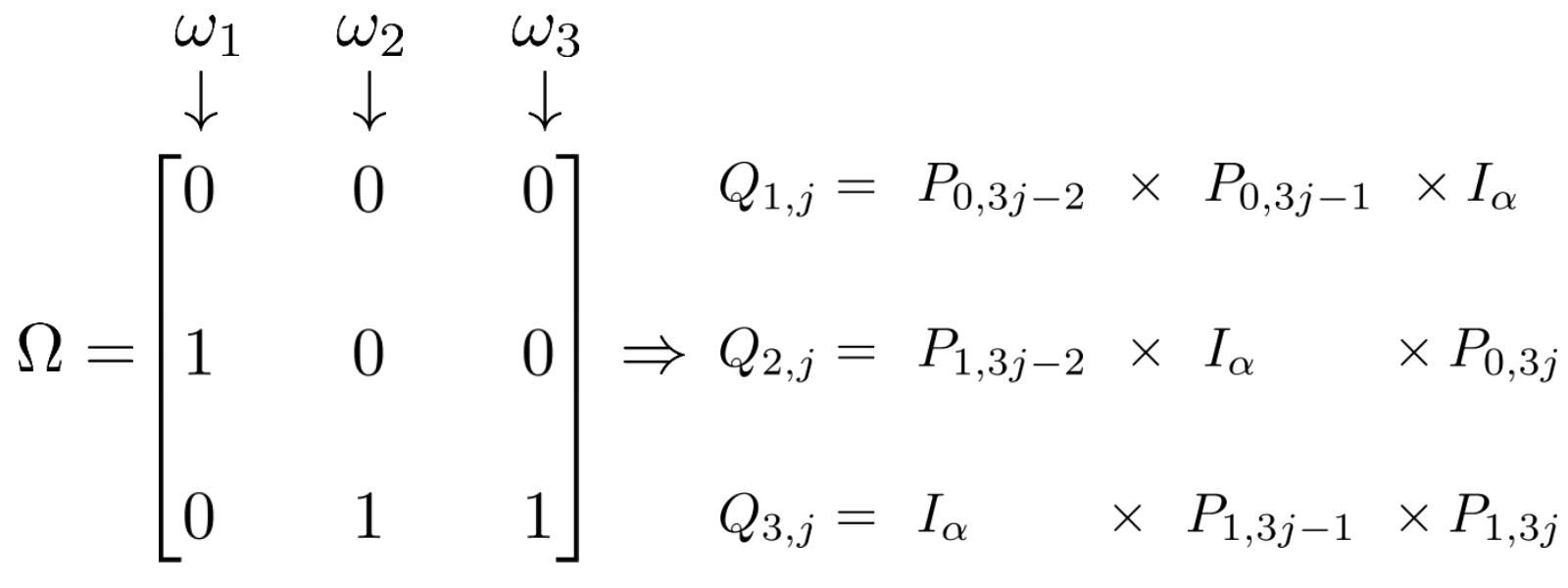}
\caption{Relation between $\omega_a$ and (\ref{eq:example_2_construction}).  }
\label{fig:example_2}
\end{figure}

\begin{IEEEproof}
Let us first explain the role of $\omega_a$ by revisiting Example \ref{example:k+3_k_k+1} via Figure \ref{fig:example_2}. Here we assumed that 
\begin{align*}
a=1 \rightarrow \hspace{1ex }\mathcal{R}_1=\{{\bf x}_{k+1},{\bf x}_{k+2}\}\hspace{1ex}\rightarrow \hspace{1ex}\omega_1 = (0,1,0)^t,\\
a=2 \rightarrow \hspace{1ex }\mathcal{R}_2=\{{\bf x}_{k+1},{\bf x}_{k+3}\}\hspace{1ex}\rightarrow \hspace{1ex}\omega_2 = (0,0,1)^t,\\
a=3 \rightarrow \hspace{1ex }\mathcal{R}_3=\{{\bf x}_{k+2},{\bf x}_{k+3}\}\hspace{1ex}\rightarrow \hspace{1ex}\omega_3 = (0,0,1)^t.
\end{align*}
In general, the matrix $\Omega=\{\omega_1 | \omega_2 | \cdots | \omega_{\binom{r}{\rho}}\}$ is designed in a way that for any choice of $a\in [\binom{r}{\rho}]$ we can always find a column in $\Omega$, denoted by $\omega_a$, such that its intersection with $r'$ rows associated with scenario $a$, forms $\{0,1,\cdots, \rho-1\}$. 

Now assume that node $i$ is failed and we are to perform an optimal systematic repair given parity repairs in $\mathcal{R}_a=\{{\bf x}_{k+d_1^{(a)}},{\bf x}_{k+d_2^{(a)}},\cdots,{\bf x}_{k+d_{\rho}^{(a)}}\}$. It is now clear that if we selected our subspaces as $S_{j,a}^i \triangleq S_{a}^i \triangleq Y_{i,a}=\{x|x \cdot e_{a+(i-1)\binom{r}{\rho}}=0 \}$, then 
\begin{align*}
&Y_{i,a}Q_{d_1^{(a)},i} = &\{x|x \cdot e_{a+(i-1)\binom{r}{\rho}}=0 \},\hspace{4ex}\\
&Y_{i,a}Q_{d_2^{(a)},i} = &\{x|x \cdot e_{a+(i-1)\binom{r}{\rho}}=1 \},\hspace{4ex}\\
&\hspace{4ex}\vdots\\
&Y_{i,a}Q_{d_{\rho}^{(a)},i} = &\{x|x \cdot e_{a+(i-1)\binom{r}{\rho}}=\rho-1 \},
\end{align*}
and hence,
\begin{align*}
\mathsf{rank}\left( \left( \begin{array}{c}
S_a^iQ_{d_1^{(a)},i}	\\
S_a^iQ_{d_2^{(a)},i}	\\
\vdots\\
S_a^iQ_{d_{\rho}^{(a)},i}
\end{array} \right) \right) = \rho\times 
\mathsf{rank}\left( Y_{i,a}\right) =\rho \frac{\alpha}{\rho} = \alpha.
\end{align*}
The second condition in Lemma \ref{lemma:interference alignment_general} is also automatically satisfied since 
\begin{align*}
Y_{i,a} \simeq Y_{i,a}Q_{1,i'} \simeq Y_{i,a}Q_{2,i'} \simeq \cdots \simeq Y_{i,a}Q_{\binom{r}{\rho},i'} \hspace{2ex} \text{for } i'\neq i.
\end{align*}
\end{IEEEproof}

Lastly, we note that optimizing the sub-packetization parameter, $\alpha$, is not the main concern. Although Construction ~\ref{cons:fixed_d} suggests a fairly large value, i.e. $\alpha=\rho^{k\binom{r}{\rho}}$, but it is clear that we do not need $\binom{r}{\rho}$ many columns in $\Omega$ to cover all the $\binom{r}{\rho}$ helper set selection scenarios. Indeed, $\alpha$ in Example \ref{example:k+3_k_k+1} can be reduced to $2^{2k}$, where $\Omega_{\text{new}} = \{\omega_1|\omega_2\}$. We leave the optimizations of this kind to future work. 

\section{Code Construction for any Helper Set}\label{sec:any_helper_nodes}

In this section, we show that Construction \ref{cons:fixed_d} in fact holds, even when an arbitrary set of $d$ helper nodes is allowed to be chosen from the $(n-1)$ surviving nodes. This generality merely imposes some additional constraints on the selection of the scaling coefficients $\lambda_{i,j}$ of the encoding matrices $A_{i,j} = \lambda_{i,j}Q_{i,j}$, where $Q_{i,j}$ is the (product) permutation matrix corresponding to $A_{i,j}$, as defined in Construction \ref{cons:fixed_d}. We now arrive at the main theorem.

\begin{theorem}{\label{thm:any_d}} {\em Construction \ref{cons:fixed_d}} gives an $[n,k,d]$ systematic--repair MSR code for any set of $d$ helper nodes, for a large enough field size for the scaling coefficients $\lambda_{i,j}$ for the encoding matrices $A_{i,j}$.
\end{theorem}

\begin{IEEEproof}\emph{ Part 1:} First, we illustrate the proof by fixing $d = k+1$, and taking an example set of helper nodes for an example failure of node ${\bf x}_1$ (or node $1$). Let us denote the (indices of the) helper set by ${\cal D}$, and let ${\cal D} = \{h,h+1,\ldots,k,k+1,\ldots,k+h\}$,
that is, there are $h$ parity nodes and $d-h = k+1-h$ systematic nodes in the helper set. Let $S_i^j({\cal D}){\bf x}_i$ denote the repair information
that node $i$ sends to help in the repair of node $j$ when ${\cal D}$ is the set of helper nodes. (Wherever clear, we ignore the ${\cal D}$ in the notation and simply write $S_i^j$.) When node $1$ fails, the information we therefore have at its replacement node can be written as:
\begin{eqnarray}\label{eq:repl-node-1}
\begin{blockarray}{cccccccc}
{\color{gray}{\bf x}_1} & {\color{gray}{\bf x}_2} & {\color{gray}\hspace{-4ex}\cdots} &\hspace{-2ex} {\color{gray}{\bf x}_h} & {\color{gray}\hspace{-2ex}{\bf x}_{h+1}} & {\color{gray}\hspace{-2ex}\cdots} & {\color{gray}\hspace{-2ex}{\bf x}_{k-1}} & {\color{gray}\hspace{-4ex}{\bf x}_k}\\
    \begin{block}{(cccccccc)}
& & & \hspace{-2ex}S_h^1 & & & &\\
& & & & \hspace{-2ex}S_{h+1}^1 & & &\\
& & & & & \hspace{-2ex}\ddots & &\\
& & & & & & \hspace{-2ex}S_{k-1}^1&\\
& & & & & & &\hspace{-4ex}S_k^1\\
S_{k+1}^1A_{11}&\hspace{-1.2ex}S_{k+1}^1A_{12}&&\hspace{-2ex}\cdots&&\hspace{-2ex}\cdots&&\hspace{-4ex}S_{k+1}^1A_{1k}\\
\vdots&\vdots&&\hspace{-2ex}\vdots&&\hspace{-2ex}\vdots&&\hspace{-4ex}\vdots\\
S_{k+h}^1A_{h1}&\hspace{-1.2ex}S_{k+h}^1A_{h2}&&\hspace{-2ex}\cdots&&\hspace{-2ex}\cdots&&\hspace{-4ex}S_{k+h}^1A_{hk}\\
    \end{block}
  \end{blockarray}
\hspace{-.7ex}\left( \hspace{-1.3ex} \begin{array}{c}
  {\bf x}_1 \\ {\bf x}_2  \\\vdots\\{\bf x}_h \\ {\bf x}_{h+1}\\\vdots \\ {\bf x}_{k-1}\\{\bf x}_k
  \end{array}\hspace{-1.8ex}\right)\hspace{-1.4ex}
\end{eqnarray}

Suppose all $S_j^1$'s in (\ref{eq:repl-node-1}) be replaced by a repair subspace $S^1$ (corresponding to Lemma \ref{lemma:interference alignment_n-1}(b)) that we would have used if ${\cal D} = \{2, 3,\ldots, k, k+1, k+2\}$. Specifically, suppose $S^1A_{1,1}$ and $S^1$ complete the space $\mathbb{F}_q^{\alpha}$.
Since $S^1$ and $S^1A_{i,j}$ denote the same subspace,
for $j \neq 1$, the components of ${\bf x}_i, i \in \{h, h+1,\ldots,k\}$ can be easily subtracted from the information coming from the parity nodes
, using that coming from the systematic nodes $h$ to $k$. Thus, in order to recover ${\bf x}_1$, we can concentrate on the following information at the replacement node:
\begin{eqnarray}\label{eq:repl-node-1b}
\begin{blockarray}{ccccc}
{\color{gray}{\bf x}_1} & {\color{gray}{\bf x}_2} & {\color{gray}\cdots} & {\color{gray}{\bf x}_{h-2}} & {\color{gray}{\bf x}_{h-1}}\\
 \begin{block}{(ccccc)}
S^1A_{11}&S^1A_{12}&\cdots&S^1A_{1,h-2}&S^1A_{1,h-1}\\
S^1A_{21}&S^1A_{22}&\cdots&S^1A_{2,h-2}&S^1A_{2,h-1}\\
\vdots&\vdots&\vdots&\vdots&\vdots\\
S^1A_{h1}&S^1A_{h2}&\cdots&S^1A_{h,h-2}&S^1A_{h,h-1}\\
    \end{block}
  \end{blockarray}
\hspace{-.5ex}\left( \hspace{-1.2ex} \begin{array}{c}
  {\bf x}_1 \\ {\bf x}_2  \\\vdots\\{\bf x}_{h-2} \\ {\bf x}_{h-1}
  \end{array}\hspace{-1ex}\right)\hspace{-2ex}
\end{eqnarray}

Let $S^1A_{i,j} = \lambda_{i,j}S^1Q_{i,j} = \lambda_{i,j}\widetilde{Q}_{i,j}S^1$, where $\widetilde{Q}_{i,j}$ is an $\alpha/2 \times \alpha/2$ matrix, and $(i,j) \neq (1,1)$. It must be noted that not only is $S^1$ dependent on the choice of ${\cal D}$, but so in turn is $\widetilde{Q}_{i,j}$. Let us also denote $S^1{\bf x}_i$ by $\widetilde{{\bf x}}_i$. Then, (\ref{eq:repl-node-1b}) can be rewritten as:
\begin{eqnarray}\label{eq:repl-node-1c}
\left(\begin{array}{cccc}
\lambda_{1,1}S^1Q_{1,1}&\hspace{-1.2ex}\lambda_{1,2}\widetilde{Q}_{1,2}&\hspace{0ex}\cdots&\hspace{-.8ex}\lambda_{1,h-1}\widetilde{Q}_{1,h-1}\\
\lambda_{2,1}\widetilde{Q}_{2,1}S^1&\hspace{-1.2ex}\lambda_{2,2}\widetilde{Q}_{2,2}&\hspace{0ex}\cdots&\hspace{-.8ex}\lambda_{2,h-1}\widetilde{Q}_{2,h-1}\\
\vdots&\hspace{-1.2ex}\vdots&\hspace{0ex}\ddots&\hspace{-.8ex}\vdots\\
\lambda_{h,1}\widetilde{Q}_{h,1}S^1&\hspace{-1.2ex}\lambda_{h,2}\widetilde{Q}_{h,2}&\hspace{0ex}\cdots&\hspace{-.8ex}\lambda_{h,h-1}\widetilde{Q}_{h,h-1}\\
\end{array}\right)
\hspace{-.5ex}\left( \hspace{-1.1ex} \begin{array}{c}
  {\bf x}_1 \\ \widetilde{{\bf x}}_2  \\\vdots\\ \widetilde{{\bf x}}_{h-1}
  \end{array}\hspace{-1.2ex}\right).\hspace{-2ex}
\end{eqnarray}

The matrix in (\ref{eq:repl-node-1c}) --- call it $M$ --- is a square matrix of dimensions $h\alpha/2 \times h\alpha/2$. A sufficient condition to recover ${\bf x}_1$ is that $M$ is invertible. Notice that the determinant of $M$, ${\sf det}(M)$, is a polynomial in the following variables: $\lambda_{i,j}, i \in [h], j \in [h-1]$. 
Hence, ${\sf det}(M)$ is a nonzero polynomial of degree $h\alpha/2$ in the given variables.
From Schwartz--Zippel--DeMillo--Lipton lemma,
 if the finite field $\mathbb{F}_q$ over which the determinant is defined has cardinality $|\mathbb{F}_q| = q > h\alpha/2$, there exist $\lambda_{i,j}$'s for which the determinant ${\sf det}(M)$ above is nonzero.

\smallskip
\emph{ Part 2}: Notice that $M$ above is defined for a particular example scenario. In general, let the number of helper nodes be $d$, the failed systematic node be $f \in [k]$, the set of helper nodes by ${\cal D} \subseteq [n]\backslash \{f\}$, the set of systematic helper nodes be ${\cal D}_s \subseteq [k]\backslash \{f\}$, and the set of parity helper nodes be ${\cal D}_p \subseteq [k+1:k+r]$. Let the number of parity helper nodes be denoted by $h$, where $h$ ranges from $d-k+1$ to $r$. Let us represent by ${\cal H}_p$ the set of parity helper nodes but indexed within $[r]$, where $i$ corresponds to node $k+i$ of the system, that is, ${\cal H}_p = \{i \,|\, k+i \in {\cal D}_p\} \subseteq [r]$. 

The matrix $M$ in (\ref{eq:repl-node-1c}), in general, can be seen to be a square matrix of dimensions $h\alpha/(d-k+1) \times h\alpha/(d-k+1)$. In particular, $M$ is a function of $f$, ${\cal D}_s$, and ${\cal H}_p$, and the determinant polynomial has degree which is a function of $|{\cal D}_p| = h$ and $d$. For each $f$, ${\cal D}_s$ and ${\cal H}_p$, we obtain a sufficiency condition that the corresponding $M$ is invertible. Therefore, the product of the corresponding determinant polynomials is a nonzero polynomial of degree
\vspace{0ex}\begin{eqnarray*}
q_{\text{ANY}} &=& k\left(\sum_{h=d-k+1}^r {r \choose h}{k-1 \choose d-h} \frac{h\alpha}{d-k+1}\right)\\
&=& \left(\sum_{h=d-k+1}^r h{r \choose h}{k-1 \choose d-h}\right)\frac{k\alpha}{d-k+1};
\end{eqnarray*}
\vspace{.1ex}consequently, there exist $\lambda_{i,j}$'s in $\mathbb{F}$ such that any systematic node is repairable with optimal repair bandwidth using any arbitrary set of $d$ helper nodes, as long as the field size $|\mathbb{F}| > q_{\text{ANY}}$.

\emph{ Part 3:} Finally, using Lemma \ref{lemma:MDS property1}, Lemma \ref{lemma:MDS property2}, and Lemma \ref{thm:fixed_d}, we obtain an $[n,k,d]$ systematic--repair MSR code for any set of $d$ helper nodes, when the field size $q > q_{\text{ANY}} + q_{\text{MDS}}$.
\end{IEEEproof}

\section{Conclusion}\label{sec:conclusion}
In this paper we presented a new construction for systematic--repair MSR codes for all possible values of parameters $[n,k,d]$. 

A more generalized construction, where a single $[n,k]$ code simultaneously satisfies the optimal repair for all $d\in\{k+1,\cdots,n-1\}$ will be introduced in a sequel paper. It is to be noted that both these generalizations come at the cost of increasing $\alpha$. A lower bound on $\alpha$ is proved in \cite{GTC14} when $d=n-1$. Whether similar bounds exist for general $[n,k,d]$ or not is left for  future work. So is the question of constructing MSR codes that also optimally repair parity nodes.

\bibliographystyle{IEEEtran}
\bibliography{references}

\begin{appendices}
\section{Interference Alignment}\label{app:lemmaIA}
\begin{IEEEproof}[Proof of Lemma \ref{lemma:interference alignment_n-1}]
We prove the result for the failure of systematic node $i = 1$. The argument generalizes for the failure of other systematic nodes. Let us assume that node ${\bf x}_1$ fails, and let each of the remaining $d = n-1$ nodes send $\beta = \alpha/r$ symbols to recover ${\bf x}_1$. In other words, node ${\bf x}_j$ (where $j \in [n]$, $j \neq i$) sends $S_j^1{\bf x}_j$ for some $\alpha/r \times \alpha$ matrix $S_j^1$. We therefore need to recover ${\bf x}_1$ from the following functions of ${\bf x}_i, i \in [k]$:

\begin{eqnarray}\label{eq:replacement-node}
\begin{blockarray}{ccccc}
{\color{gray}{\bf x}_1} & {\color{gray}{\bf x}_2} & {\color{gray}{\bf x}_3} & {\color{gray}\cdots} & {\color{gray}{\bf x}_{k}}\\[0.5ex]
    \begin{block}{(ccccc)}
    & S_2^1 & & & \\
    & & S_3^1 & & \\
    & & & \ddots &\\
    & & & & S_k^1\\
    S_{k+1}^1A_{1,1} & S_{k+1}^1A_{1,2} &\cdots & \cdots & S_{k+1}^1A_{1,k}\\[0.5ex]
    S_{k+2}^1A_{2,1} & S_{k+2}^1A_{2,2} &\cdots & \cdots & S_{k+2}^1A_{2,k}\\
   \vdots & \vdots & \vdots & \vdots & \vdots \\
    S_{k+r}^1A_{r,1} & S_{k+r}^1A_{r,2} &\cdots & \cdots & S_{k+r}^1A_{r,k}\\
    \end{block}
  \end{blockarray}
\left(\hspace{-0.5ex}  \begin{array}{c}
  {\bf x}_1 \\ {\bf x}_2  \\ {\bf x}_3 \\ \vdots \\ {\bf x}_{k}
  \end{array}\hspace{-0.5ex}\right).
\end{eqnarray}

{\em Necessity}: Suppose the systematic vectors ${\bf x}_2$ through ${\bf x}_k$ be the zero vectors. Then, (\ref{eq:replacement-node}) simplifies to:
\begin{eqnarray*}
\left(\begin{array}{c}
 S_{k+1}^1A_{1,1}\\[0.5ex]
    S_{k+2}^1A_{2,1} \\
   \vdots\\
    S_{k+r}^1A_{r,1}
\end{array}\right){\bf x}_1,
\end{eqnarray*}
where the matrix is an $\alpha \times \alpha$ square matrix. Since ${\bf x}_1$ is recoverable, it is necessary that the matrix be non-singular, thus proving the signal recovery conditions (\ref{eq:signal-recovery}) and (\ref{eq:signal-recovery2}). Note that this also implies that all encoding matrices $A_{i,j}, i \in [r], j \in [k]$, are non-singular. s

Suppose now, without loss of generality, that the interference alignment condition (\ref{eq:IAcondition}) is not satisfied for $j = 2$. Again, without loss of generality, let $S_2^1 \not \backsimeq S_{k+1}^1A_{1,2}$. This implies that 
\begin{eqnarray}\label{eq:contra}
\mathsf{rank}\left(\left(\begin{array}{c}S_2^1\\S_{k+1}^1A_{1,2}\end{array}\right)\right) &=& \frac{\alpha}{r}+\epsilon,
\end{eqnarray}
for some $\epsilon > 0$. Since ${\bf x}_1$ is recoverable, from (\ref{eq:replacement-node}), we have access to the following information at the replacement node:
\begin{eqnarray}\label{eq:replacement-node2}
\begin{blockarray}{ccccc}
{\color{gray}{\bf x}_1} & {\color{gray}{\bf x}_2} & {\color{gray}{\bf x}_3} & {\color{gray}\cdots} & {\color{gray}{\bf x}_{k}}\\[0.5ex]
    \begin{block}{(ccccc)}\hspace{1ex}
    I_{\alpha} & & & &\\
    & S_2^1 & & & \\
    & & S_3^1 & & \\
    & & & \ddots &\\
    & & & & S_k^1\\
0 & S_{k+1}^1A_{1,2} &\cdots & \cdots & S_{k+1}^1A_{1,k}\hspace{0.5ex}\\
    \end{block}
  \end{blockarray}
\left(\hspace{-0.5ex}  \begin{array}{c}
  {\bf x}_1 \\ {\bf x}_2  \\ {\bf x}_3 \\ \vdots \\ {\bf x}_{k}
  \end{array}\hspace{-0.5ex}\right).
\end{eqnarray}
From (\ref{eq:contra}), the rank of the matrix in (\ref{eq:replacement-node2}) is at least
\begin{eqnarray*}
\alpha + \frac{\alpha}{r}+\epsilon + (k-2)\frac{\alpha}{r} &>& (n-1)\frac{\alpha}{r},
\end{eqnarray*}
the total number of symbols available at the replacement node. In other words, we are able to recover more number of linearly independent symbols that are functions of the systematic data vectors ${\bf x}_1$ through ${\bf x}_k$, than the number of repair symbols available at the replacement node --- a contradiction! Thus, conditions (\ref{eq:IAcondition}) and (\ref{eq:IAcondition2}) must be true.

\smallskip
{\em Sufficiency}: Suppose that we have the required repair matrices $S_j^1$ that satisfy the signal recovery and interference alignment conditions (\ref{eq:signal-recovery}) and (\ref{eq:IAcondition}). Using (\ref{eq:IAcondition}), we can eliminate the contribution of systematic vectors ${\bf x}_2$ through ${\bf x}_k$ in the information transmitted by the parity nodes (that is, the last $r$ rows in (\ref{eq:replacement-node})). For instance, $S_2^1 \backsimeq S_{k+1}^1A_{1,2}$ implies that $S_{k+1}^1A_{1,2} = BS_2^1$, for some $\alpha/r \times \alpha/r$ matrix $B$, and therefore the contribution of $S_{k+1}^1A_{1,2}{\bf x}_2$ can be removed from the repair information transmitted by the parity node $k+1$ using the repair information $S_2^1{\bf x}_2$ (or equivalently, $BS_2^1{\bf x}_2$) transmitted by systematic node $2$. Using (\ref{eq:signal-recovery}), it is then easy to recover ${\bf x}_1$.
\end{IEEEproof}

\end{appendices}

\end{document}